\documentclass[sigconf]{acmart}
\usepackage{amsmath,amssymb,amsfonts}
\usepackage{algorithmic}
\usepackage{graphicx}
\usepackage{textcomp}
\usepackage{xcolor}
\usepackage[euler]{textgreek}
\usepackage{color}
\usepackage{url}
\usepackage{todonotes}
\usepackage{multirow}
\usepackage{tabularx}
\usepackage{makecell}

\def\BibTeX{{\rm B\kern-.05em{\sc i\kern-.025em b}\kern-.08emT\kern-.1667em\lower.7ex\hbox{E}\kern-.125emX}}

\copyrightyear{2019}
\acmYear{2019}
\setcopyright{acmlicensed}
\acmConference[UCC '19]{UCC '19: IEEE/ACM International Conference on Utility and Cloud Computing Doctoral Symposium}{December 02--05, 2019}{Auckland, New Zealand}
\acmBooktitle{UCC '19: IEEE/ACM International Conference on Utility and Cloud Computing Doctoral Symposium, December 02--05, 2019, Auckland, New Zealand}
\acmPrice{15.00}
\acmDOI{10.1145/1122445.1122456}
\acmISBN{978-1-4503-9999-9/18/06}

\begin{document}

\title{Performance Estimation of Container-Based\\ Cloud-to-Fog Offloading}

\author{Ayesha Abdul Majeed, Peter Kilpatrick, Ivor Spence, and Blesson Varghese}
\authornote{Corresponding Author}
\email{{aabdulmajeed01, p.kilpatrick, i.spence, b.varghese}@qub.ac.uk}
\affiliation{%
  \institution{Queen's University Belfast, UK}
}

\renewcommand{\shortauthors}{Author1 et al.}

\begin{abstract}
Fog computing offloads latency critical application services running on the Cloud in close proximity to end-user devices onto resources located at the edge of the network.
The research in this paper is motivated towards characterising and estimating the time taken to offload a service using containers, which is investigated in the context of the `Save and Load' container migration technique. 
To this end, the research addresses questions such as whether fog offloading can be accurately modelled and which system and network related parameters influence offloading.
These are addressed by exploring a catalogue of 21 different metrics both at the system and process levels that is used as input to four estimation techniques using collective model and individual models to predict the time taken for offloading. 
The study is pursued by collecting over 1.1 million data points and the preliminary results indicate that offloading can be modelled accurately. 
\end{abstract}

\ccsdesc[500]{Computing methodologies~Distributed computing methodologies}
\ccsdesc[300]{Computer systems organization~Embedded and cyber-physical systems}

\keywords{Fog computing, containers, offloading, Edge computing}

\maketitle

\section{Introduction}
\label{sec:introduction}
\begin{figure}[t]
    \centering
    \includegraphics[width=0.4\textwidth]{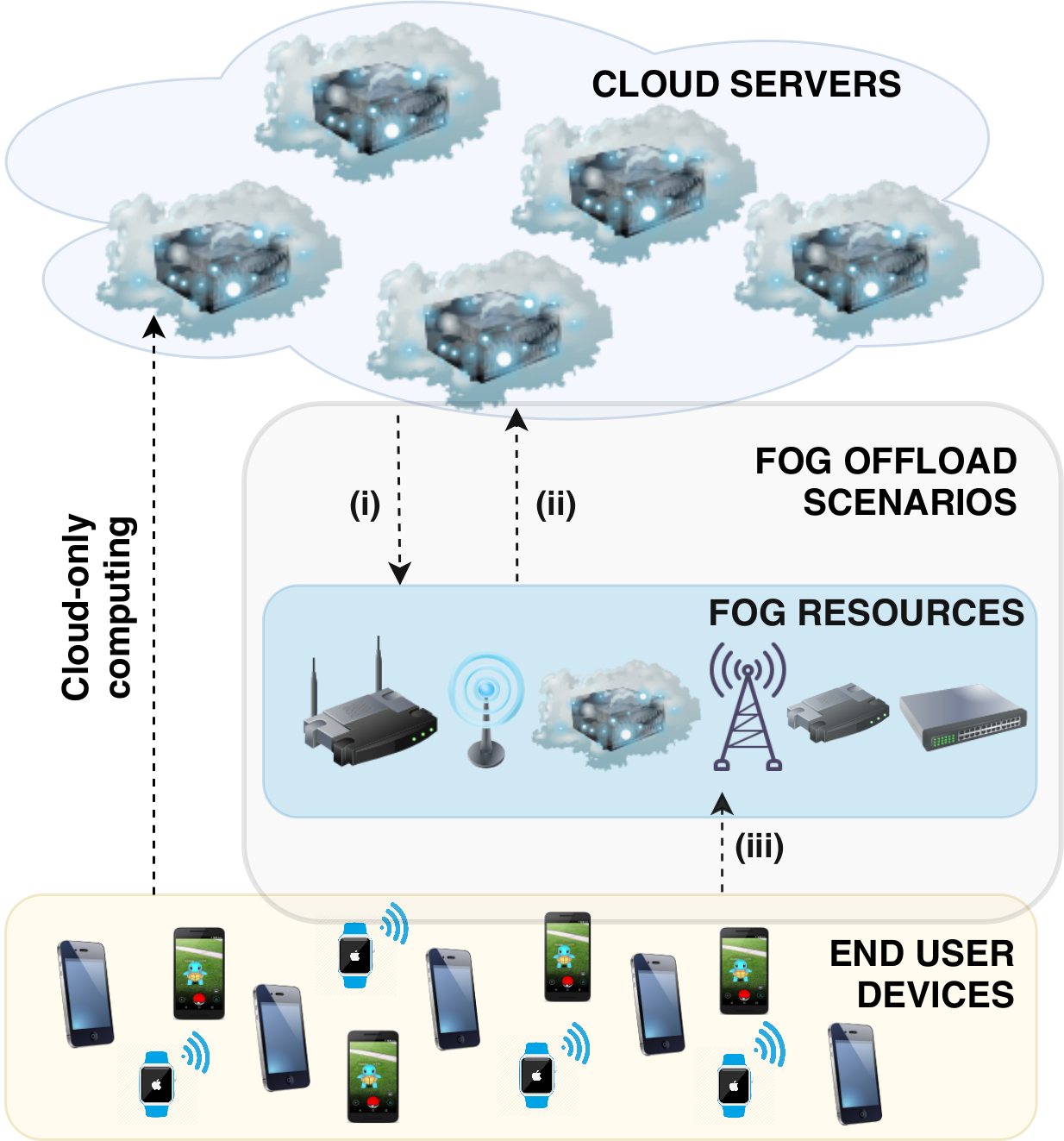}
    \caption{Potential Fog offload scenarios, namely (i) Cloud-to-Fog, (ii) Fog-to-Cloud, and (iii) Device-to-Fog}
    \label{fig:offloading}
\end{figure}

Offloading services from end user devices to a Cloud data centre in a Cloud-only computing model was developed to meet the computational and energy requirements of hardware limited and battery powered devices. Since then Fog computing has evolved in which resources located on the edge of the network, such as micro data centres or compute-enabled routers, are leveraged to make applications more responsive. 
Three scenarios as illustrated in Figure~\ref{fig:offloading} are envisioned when offloading services to or from the Fog: (i) offloading from the Cloud to the Fog, which is closer to end-user devices for minimising communication latencies~\cite{satyanarayanan2017emergence}, (ii) offloading from the Fog to the Cloud to meet additional compute requirements~\cite{management2018}, and (iii) offloading from user devices to the Fog to satisfy compute requirements unavailable on the devices~\cite{osmoticcomputing-01}. This paper focuses on the Cloud-to-Fog offload scenario. 


Although offloading is reported to be beneficial~\cite{satyanarayanan2017emergence}, applications can be adversely affected if there are large offloading overheads that result in application downtime when a service is offloaded~\cite{ma2017efficient}. This naturally affects user experience. The time taken to offload a service is an indicator of the downtime of an application. If there are multiple methods to offload a service, then it would be beneficial to estimate the offload time taken by each offloading method so that the down time can be estimated. This estimation by useful in the following use-cases: 

(i) \textit{Automated Fog software development environments}: to provide application developers estimates of deployment time and down time of services as they design and develop Fog applications. Rather than having to empirically test each application, which is time-consuming and cumbersome, predictive models can provide insight into feasibility of offloading given a set of environment conditions.

(ii) \textit{Adaptive decision-making in Fog orchestration platforms}: to support decision-making in orchestration platforms where it may be possible to offload multiple combination of services in a transient environment using different techniques, predictive models can make decisions on which offloading technique is best suited given the current state of the system and associated requirements.

(iii) \textit{Simulation platforms}: predictive models that can provide (near) accurate estimates of the time taken to offload a service from the Cloud to the Fog or from an edge device to the Fog can be useful in simulation platforms that are used by researchers or practitioners that do not have direct access to the physical infrastructure for developing their applications.

This paper investigates one container-based offloading technique, namely Save and Load, in the context of offloading from the Cloud to the Fog. The aim is to characterise and estimate the time taken for offloading a service. For this a catalogue of 21 metrics that is relevant to the entire Cloud and Fog system and the offloading process is considered. Two estimation methods are explored, namely using a collective model and a set of individual models to estimate the offload time using four machine learning based predictive models. The models are built on over 1.1 million data points obtained from two experimental platforms.
Preliminary results indicate that the time taken to offload can be estimated with reasonable accuracy and the method using individual models are more accurate than using a collective model.

The remainder of this paper is organised as follows. 
Section~\ref{sec:techniques} discusses the scenarios and approaches for offloading.
Section~\ref{sec:estimationmodels} proposes two methods for estimating the performance of offloading and highlights the models for estimating performance.
Section~\ref{sec:experiments} presents the experimental studies pursued on two different platforms. 
Section~\ref{sec:relatedwork} highlights research that is relevant to the discussion of this paper. 
Section~\ref{sec:conclusions} concludes this paper by considering future work.

\section{Container-based Offloading}
\label{sec:techniques}
This section presents the three offloading scenarios highlighted in Figure~\ref{fig:offloading}, the key approaches for offloading, and focuses on one such approach that is based on containers, namely Save and Load.
 
\subsection{Offloading Scenarios} 
 
Fog offloading can be considered in the following three scenarios, which are illustrated in Figure~\ref{fig:offloading}: 

(i) \textit{Cloud-to-Fog offloading}~\cite{wang2017enorm, chen2018predictive, pham2016towards} - this refers to transferring services of an application from a Cloud resource on to a Fog resource to meet latency and ingress bandwidth demands. Since the offloaded service is closer to the device that generates data, it reduces the communication latency and (pre)-processes data closer to the source, thereby reducing the volume of data that needs to be transferred to the cloud. 

(ii) \textit{Fog-to-Cloud offloading}~\cite{wang2017enorm,enguehard2018popularity} - this refers to transferring the services that are resident on a Fog resource to the Cloud. This may be done due to a change in the life-cycle of the application service - the service on the Fog resource needs to be terminated and resumed on the Cloud from where it was originally offloaded. Alternatively, this may be because the service requires additional resources (CPU cores, storage, memory) that are not available on the Fog resource, but can only be satisfied on the Cloud. 

(iii) \textit{Device-to-Fog offloading}~\cite{osmoticcomputing-01,yousefpour2018reducing, yu2017novel} - this refers to transferring services of an application from one or a collection of end user devices or sensors to a Fog resource. This is done in order to preserve battery life of devices or to meet the computational demands of workloads that cannot be executed on user devices due to limited form factor and weak processing capabilities. The offloaded service executes on the Fog resource and the resulting output is provided back to the devices or sensors. 

The focus of this paper is the Cloud-to-Fog offloading scenario.

\subsection{Offloading Approaches}
There are two dominant approaches that facilitate offloading. The first is Virtual Machine (VM) migration. VM hand-off is one approach that has been explored to move a service from one resource to another to support user mobility in the context of cloudlets~\cite{ha2015adaptive}. A synthesis technique is adopted in which the VM is divided into two stacked overlays so as to optimise the downtime when VM hand-off occurs. 

A second approach to facilitate offloading is container migration. Containers are an alternate virtualisation approach that are lightweight and portable, thereby making offloading quicker than using VMs~\cite{soltesz2007container}. Therefore, containers are popularly investigated in the context of Fog computing, in which Fog resources have limited resources when compared to the Cloud and therefore require more lightweight and portable virtualisation technologies~\cite{wang2017enorm,management2018}. Popular container technologies, include LXC\footnote{\url{https://linuxcontainers.org/}} and Docker\footnote{\url{https://www.docker.com/}} and they support migration, which is required for offloading. 

\subsubsection*{Container-Based Offloading}
\begin{figure}[ht]
    \hspace{-25pt}
    \includegraphics[width=0.5\textwidth]{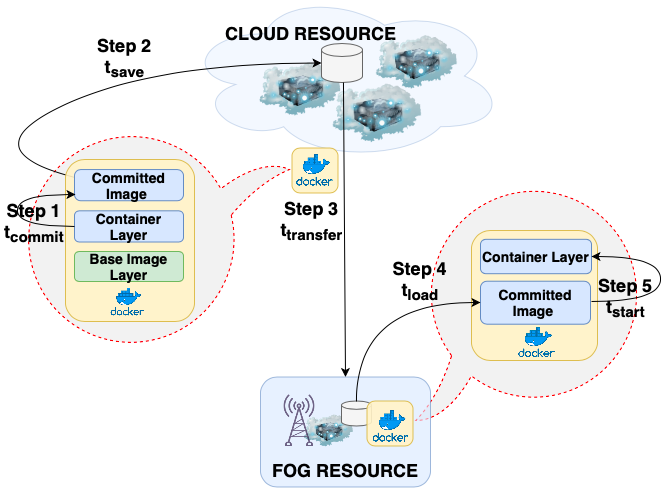}
    \caption{Five steps in offloading a container from the Cloud to the Fog using the Save and Load migration technique}
    \label{fig:dockercontaineroffloading}
\end{figure}

Fog offloading, in this paper, is explored in the context of Docker. Docker packages an application in the form of images that consist of a file system with the required libraries, executables and configuration files. In practice, a Docker image comprises a series of layers that is stacked on top of a base image, for example, Ubuntu operating system. When a container is executed Docker mounts all the layers of the image as `read-only' using the Union File System (UnionFS) and the top layer as a writable layer as shown in Figure~\ref{fig:dockercontaineroffloading}.

Docker supports two migration techniques: (i) stateful - the state of the running application is transferred with the container image, and (ii) stateless - the state of the application is not transferred, instead a separate instance of the container is run elsewhere without its previous state. 
Stateful migration can be achieved by using CRIU (Checkpoint/Restore in Userspace)\footnote{\url{https://criu.org/Main_Page}} approach. Stateless migration techniques include methods, such as Pull and Push, Export and Import, and Save and Load. This is a preliminary investigation into different migration techniques that can be used for offloading and therefore only one stateless migration technique, namely `\textit{Save and Load}' is considered within the scope of this paper. The other techniques will be reported elsewhere. 

\subsection{Save and Load Technique for Offloading}
Figure~\ref{fig:dockercontaineroffloading} shows the Cloud-to-Fog offloading scenario using the Save and Load migration technique (the Fog-to-Cloud scenario will execute in a similar manner with the steps that occur on the Cloud shown in the figure on the Fog and vice-versa). The goal of this technique is to transfer the image of a running container and the underlying base layers. This offloading technique is a five step method as illustrated below: 

\textit{Step 1 - Commit}: A container is instantiated when it boots up from a series of base image layers 
When there is a request for offloading from the Cloud server, the commit operation stops the running container and saves the current state of the container, the accompanying stacked layers, and the modifications that were made within the container as a new image, referred to as the committed image. This newly created image is stored in the local image registry of the Cloud server. The time taken for this step (store the configuration and run-time state of the container, create an image, and store the image in the local registry) is denoted as $t_{commit}$.

\textit{Step 2 - Save}: The save operation, converts the committed image to a compressed \texttt{.tar} file and saves it on the hard disk of the Cloud server. The compressed file contains information of the contents of the container, including its parent layers and the size of each layer. The time taken to convert the committed image to the compressed file is denoted as $t_{save}$. 

\textit{Step 3 - Transfer}: The compressed image file is transferred to the Fog resource using a network transfer protocol, such as FTP, SCP or rsync; $t_{transfer}$ time captures the time taken to transfer the compressed file to the Fog server.

\textit{Step 4 - Load}: This operation initially decompresses the \texttt{.tar} image file and loads it from the hard disk on the Fog resource as an image into local image registry. This time is captured as $t_{load}$.

\textit{Step 5 - Start}: A container from the image in the registry is booted up on the Fog resource; $t_{start}$ captures this time.

Based on the above five steps, the total time taken to offload a container-based service from the Cloud to the Fog or vice-versa is represented as:
 
\begin{equation}
t_{offload} = t_{commit} + t_{save} + t_{transfer} + t_{load} + t_{start}
\label{eqn:toffload}
\end{equation}

In the Cloud-to-Fog offload scenario,
$t_{commit}$ and $t_{save}$ are the operations on the Cloud server, and $t_{load}$ and $t_{start}$ are for the Fog resource. The transfer time will be for transferring the compressed file from the Cloud to the Fog. In the Fog-to-Cloud offload scenario, the commit and save times will be for the operation on the Fog resource, and the load and start for the operations on the Cloud server. 

In this research it is assumed that Fog applications are designed and developed as micro-services (a collection of services can be orchestrated as a workflow and each service can be deployed as a container). The advantage of this offloading technique is that the base image layers are also transferred from the source (for example, the Cloud server in the Cloud-to-Fog offload scenario) to the destination (Fog resource in the Cloud-to-Fog offload scenario). This eliminates any need for reinstallation of the underlying libraries. However, there will be a trade-off with the size of the image transferred. This benefit is also apparent when a service needs to be migrated from one Fog resource to another.

A potential advantage is when multiple containers that rely on the same underlying libraries need to be offloaded. In this case, the base image layers will not be 
duplicated for multiple container services when it is offloaded. 


\section{Estimating Offload Time}
\label{sec:estimationmodels}
In this section, the parameters that influence the Save and Load offloading technique, the methods used for estimation and the machine learning algorithms used for building the estimation model that is used for predicting ($t_{offload}$) are presented.

\subsection{Methods for Estimation}
Table~\ref{tab:tab1} describes a catalogue of parameters that are considered in this paper by the estimation models while offloading a container. Two types of parameter are considered, namely runtime and offline parameters. The \textit{runtime} parameters are collected when the five steps of the Save and Load technique are executed to offload a container. These parameters relate to both the offloading process and the entire Cloud and/or Fog system as highlighted in the table. The network properties between the Cloud and the Fog are considered to capture the state of the network when the offload occurs. The \textit{offline} parameters are statically determined and do not change during the offload process; for example number of cores, network bandwidth and container image size.

\begin{table}[t]
\caption{Parameters that impact overall Cloud-to-Fog offloading time; Bps - Bytes per second, bps - bits per second, BW - bandwidth}
\label{tab:tab1}
\vspace{-3pt}
\begin{tabular}{|l|l|c|l|}
\hline
\textbf{Parameter} & \textbf{Description} & \multicolumn{1}{l|}{\textbf{\begin{tabular}[c]{@{}c@{}}System/\\Process\end{tabular}}} & \textbf{\begin{tabular}[c]{@{}c@{}}Cloud/\\Fog\end{tabular}} \\ \hline
\multicolumn{4}{|c|}{\textit{Runtime}} \\ \hline
$P_{1}$, $P_{7}$ & CPU utilisation (\%) & \multirow{3}{*}{System} & \multirow{3}{*}{\begin{tabular}[c]{@{}l@{}}$P_{1}$ - $P_{3}$ (Cloud)\\$P_{7}$ - $P_{9}$ (Fog)\end{tabular}} \\ \cline{1-2}
$P_{2}$, $P_{8}$ & Memory utilisation (\%) &  &  \\ \cline{1-2}
$P_{3}$, $P_{9}$ & Disk utilisation (\%) &  &  \\ \hline
$P_{4}$, $P_{10}$ & CPU utilisation (\%) & \multirow{3}{*}{\begin{tabular}[c]{@{}c@{}}Offloading \\ Process\end{tabular}} & \multirow{3}{*}{\begin{tabular}[c]{@{}l@{}}$P_{4}$ - $P_{6}$ (Cloud)\\$P_{10}$ - $P_{12}$ (Fog)\end{tabular}} \\ \cline{1-2}
$P_{5}$, $P_{11}$ & Memory utilisation (\%) &  &  \\ \cline{1-2}
$P_{6}$, $P_{12}$ & Disk throughput (Bps) &  &  \\ \hline
\multicolumn{4}{|c|}{\textit{Offline}} \\ \hline
$P_{13}$ & Image size (MB) & \multicolumn{2}{c|}{\begin{tabular}[c]{@{}c@{}}Offloaded container\end{tabular}} \\ \hline
$P_{14}$, $P_{17}$ & No of Cores & \multicolumn{2}{c|}{\multirow{3}{*}{\begin{tabular}[c]{@{}c@{}}$P_{14}$ - $P_{16}$  (Cloud)\\ $P_{17}$ - $P_{19}$ (Fog)\end{tabular}}} \\ \cline{1-2}
$P_{15}$, $P_{18}$ & Memory Size (GB) & \multicolumn{2}{c|}{} \\ \cline{1-2}
$P_{16}$, $P_{19}$ & Hard disk Size (GB) & \multicolumn{2}{c|}{} \\ \hline
$P_{20}$ & Network BW (bps) & \multicolumn{2}{c|}{\multirow{2}{*}{\begin{tabular}[c]{@{}c@{}}Between \\ Cloud and Fog\end{tabular}}} \\ \cline{1-2}
$P_{21}$ & Network Latency (ms) & \multicolumn{2}{c|}{} \\ \hline
\end{tabular}
\end{table}

\begin{table}[t]
\caption{Parameters that impact the offloading sequence}
\label{tab:tab2}
\centering
\vspace{-3pt}
\begin{tabular}{| l | l |}
\hline
\multicolumn{1}{|c|}{\textbf{Time}} & \multicolumn{1}{c|}{\textbf{Parameters that impact}} \\ \hline
$t\textsubscript{commit}$ & $X_{commit} = \{$ P\textsubscript{1}, $\cdots$, P\textsubscript{6}, 
P\textsubscript{13}, $\cdots$, P\textsubscript{19} $\}$ \\
$t\textsubscript{save}$ & \begin{tabular}[c]{@{}l@{}}$X_{save} = \{$ P\textsubscript{1}, $\cdots$, P\textsubscript{19}, P\textsubscript{13}, P\textsubscript{19}\end{tabular} \}\\
$t\textsubscript{transfer}$ & $X_{transfer} = \{$ P\textsubscript{13}, P\textsubscript{20}, P\textsubscript{21} $\}$\\
$t\textsubscript{load}$ & $X_{load} = \{$ \begin{tabular}[c]{@{}l@{}}P\textsubscript{7}, $\cdots$ P\textsubscript{19} $\}$\end{tabular} \\
$t\textsubscript{start}$ & \begin{tabular}[c]{@{}l@{}}$X_{start} = \{$ P\textsubscript{7}, $\cdots$, P\textsubscript{19} $\}$\end{tabular} \\ \hline
\end{tabular}
\end{table}

Two methods are used to estimate the offload time -the first uses a single estimation model and the second uses multiple estimation models for the individual time components shown in Equation~\ref{eqn:toffload}.

The first method is \textit{using a collective model}, referred to as (CM), which is a reference to the use of a single model that estimates the offload time. The collective model uses all the parameters listed in Table~\ref{tab:tab1} as input. Consider a collective model, $M_{collective}$ that estimates the offload time and $X_{offload} = \{P_{1},\cdots, P_{21}\}$  be the input to the model, then we represent $t_{offload}$ = $M_{collective}(X_{offload})$. 

The second method is \textit{using individual models}, which refers to the use of separate models for estimating the individual times of Equation~\ref{eqn:toffload} ($t_{commit}$, $t_{save}$, $t_{transfer}$, $t_{load}$ and $t_{start}$).
Table~\ref{tab:tab2} shows the parameters that affect the individual times. Let $M_{commit}$ be an individual model to estimate $t_{commit}$ using the input $X_{commit}$ shown in Table~\ref{tab:tab2}. Then the estimation of $t_{commit}$ = $M_{commit}(X_{commit})$.
Similarly, $t_{save}$ = $M_{save}(X_{save})$, 
$t_{transfer}$ = $M_{transfer}(X_{transfer})$, 
$t_{load}$ = $M_{load}(X_{load})$, and
$t_{start}$ = $M_{start}(X_{start})$.
The offload time can be estimated as shown in Equation~\ref{eqn:individual}.
\begin{multline}
\label{eqn:individual}
    t_{offload} = M_{commit}(X_{commit})+M_{save}(X_{save})+\\
    M_{transfer}(X_{transfer})+M_{load}(X_{load})+M_{start}(X_{start})
\end{multline} 


\subsection{Models for Estimation}
\label{subsec:models}
Four machine learning algorithms were explored for predicting the offload time. The approach used for estimation is based on historical data that is collected from the experimental platform (presented in Section~\ref{sec:experiments}) to predict $t_{offload}$. The algorithms used are:

(i) Multivariate Linear Regression (MLR): The model developed using this algorithm captures the relationship between multiple input variables $X = \{ P_{1}, P_{2},\cdots, P_{n}\}$ and the dependent output variable $t_{offload}$ by a straight line equation~\cite{pham2017predicting}.

(ii) Polynomial Multivariate Regression (PMR): The model developed is a regression technique which captures the relationship between the input variables $X = \{ P_{1}, P_{2},\cdots, P_{n}\}$ and the dependent output variable $t_{offload}$ as an $n^{th}$ degree polynomial in $X$.

(iii) Random Forest Regression (RFR): An ensemble model generates $k$ different training subsets from the original data set, and then $k$ different decision trees are built based on the generated training subsets. Each sample of the testing data set is predicted by all decision trees, and the final result is obtained by averaging a score that is specific to each decision tree~\cite{pham2017predicting}.

(iv) Support Vector Regression (SVR): This estimation model uses non linear mapping to transform input data into a higher-dimensional feature space~\cite{sapankevych2009time}. 

Multiple estimation methods are explored since it is currently unknown whether using a single collective model or multiple models are most suited for predicting the offload time.

\section{Experimental Studies}
\label{sec:experiments}
This section presents the experimental setup and the preliminary results obtained from running experiments.

\textit{Experimental setup}:
The methods proposed above, namely the Collective Model (CM) and the Individual Models (IM) for estimating the performance of offloading using four estimation models, namely MLR, PMR, RFR and SVR are evaluated on two different platforms (each platform is the combination of a Cloud and Fog VM that executes the Save and Load technique). Docker 18.09-ce is installed on all VMs of the experimental platforms.

Both experimental platforms use 64-bit x86 architectures for the Cloud and Fog environment. Although Fog nodes may be hardware limited ( and using ARM-based processors) when compared to the Cloud, recent Fog-enabled nodes, such as the Dell Edge Gateway 5000, use 64-bit x86 processors~\cite{puliafito2019fog}. 

The first platform is the combination of a Cloud VM running Ubuntu 18.10 with 6 virtual CPUs, 30GB hard disk and 6GB RAM and Fog VM running Ubuntu 18.10 with 2 virtual CPUs, 20GB hard disk and 2GB RAM. 
The network bandwidth between the Cloud and Fog VMs are emulated using the Linux Traffic Control (\texttt{tc})\footnote{https://linux.die.net/man/8/tc} package. The bandwidth is varied as 25Mbps, 50Mbps, 100Mbps, 1000Mbps with a latency of 30ms. These values are chosen based on research reported in the literature that establishes these as baselines~\cite{ma2017efficient}. 

The second platform is another combination of a Cloud VM and Fog VM running OpenStack. Both VMs run Ubuntu 18.10; the Cloud VM has 4 virtual CPUs, 80GB hard disk space and 8GB virtual RAM where as the Fog VM has 2 virtual CPUs, a 40GB hard disk, and 4GB virtual RAM. The default network connection is 3.2Mbps.

A Cloud VM, a Fog VM and an observing process on the Cloud server are employed for obtaining the parameters listed in Table~\ref{tab:tab1}. The parameters collected on the Cloud VM and the Fog VM are sent to the observing process. 


During the offloading process, the values of the run-time parameters are collected at every one-second interval. System CPU utilisation is captured by $P_{1}$ and $P_{7}$, which are obtained by monitoring \texttt{/proc/stat}. CPU utilisation of the offloading process captured by $P_{4}$ and $P_{10}$ are obtained by monitoring \texttt{/proc/PID/}. RAM utilisation of the system, denoted by $P_{2}$ and $P_{8}$ are obtained by the Linux utility tool \texttt{ps}. RAM utilisation of the offloading process is monitored using \texttt{/proc/PID/smaps} file for the parameters $P_{5}$ and $P_{11}$. Disk utilisation of the system, denoted by $P_{3}$ and $P_{9}$ are obtained using \texttt{iotop} utility. Disk throughput of the offloading process is obtained by monitoring \texttt{/proc/PID/io} to record the number of bytes written to and read from disk for parameters $P_{6}$ and $P_{12}$. 

The values for offline parameters (i) $P_{13}$ is the size of the container that is offloaded, (ii) $P_{14}$ - $P_{19}$, were obtained through settings defined for the Cloud VM and Fog VM, 
and (iii) $P_{20}$ and $P_{21}$ are acquired during the network configurations set using \texttt{tc}.
 
To simulate the varying availability of CPU, memory and hard disk resources in the experimental environment, the CPU, memory and I/O stress were gradually increased for different experimental runs using \texttt{stress-ng}\footnote{https://manpages.ubuntu.com/manpages/artful/man1/stress-ng.1.html}; CPU stress was increased by 10\%, memory stress on the Cloud VM by units of 1GB until 75\% of capacity and for the Fog VM by units of 512MB until 75\% of capacity, and disk stress on the Cloud VM by units of 4GB until 75\% of capacity and for the Fog VM by units of 2GB until 75\% of capacity.

The collected data values using the estimation methods presented in Section~\ref{sec:estimationmodels} are used to build the model for estimating $t_{offload}$.

The dataset that is generated from all the parameters for the prediction model consisted of 836 instances across the two experimental platforms for the Cloud-to-Fog offload scenario for varying combinations of the offline parameters. Since the runtime parameters are collected at one-second intervals each Cloud-to-Fog offload scenario for a given set of offload parameters will generate a large number of intermediate instances (on an average 65 intermediate instances). 
The runtime values are averaged to obtain the aggregate instances. A total of $21 \times 836 \times 65 = 1,141,140$ data points are used. 

\begin{figure}[t]
    \centering
    \includegraphics[width=0.5\textwidth]{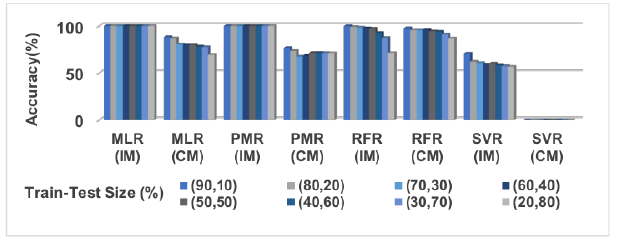}
    \caption{Accuracy of the prediction models under varying training and testing data sizes}
    \label{fig:accuracy1}
\end{figure}

\begin{figure}[t]
    \centering
    \includegraphics[width=0.5\textwidth]{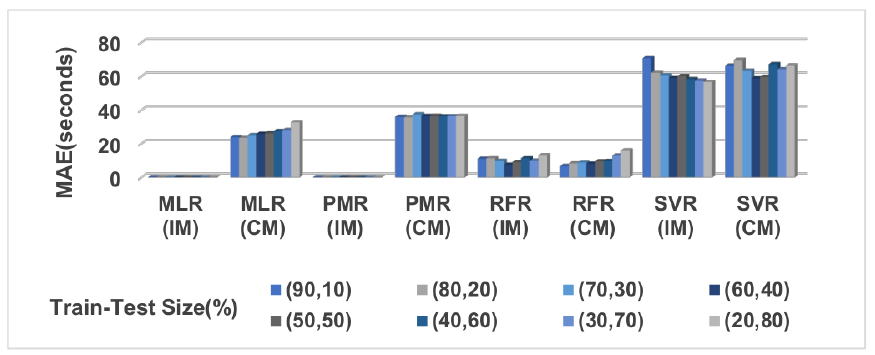}
    \caption{Mean Absolute Error (MAE) of the prediction models under varying training and testing data sizes}
    \label{fig:mae1}
\end{figure}

\begin{figure}[t]
    \centering
    \includegraphics[width=0.5\textwidth]{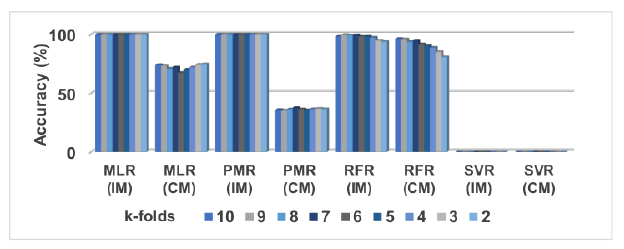}
    \caption{Accuracy of the prediction models for varying sample sizes using k-fold cross validation}
    \label{fig:accuracy2}
\end{figure}

\begin{figure}[t]
    \centering
    \includegraphics[width=0.5\textwidth]{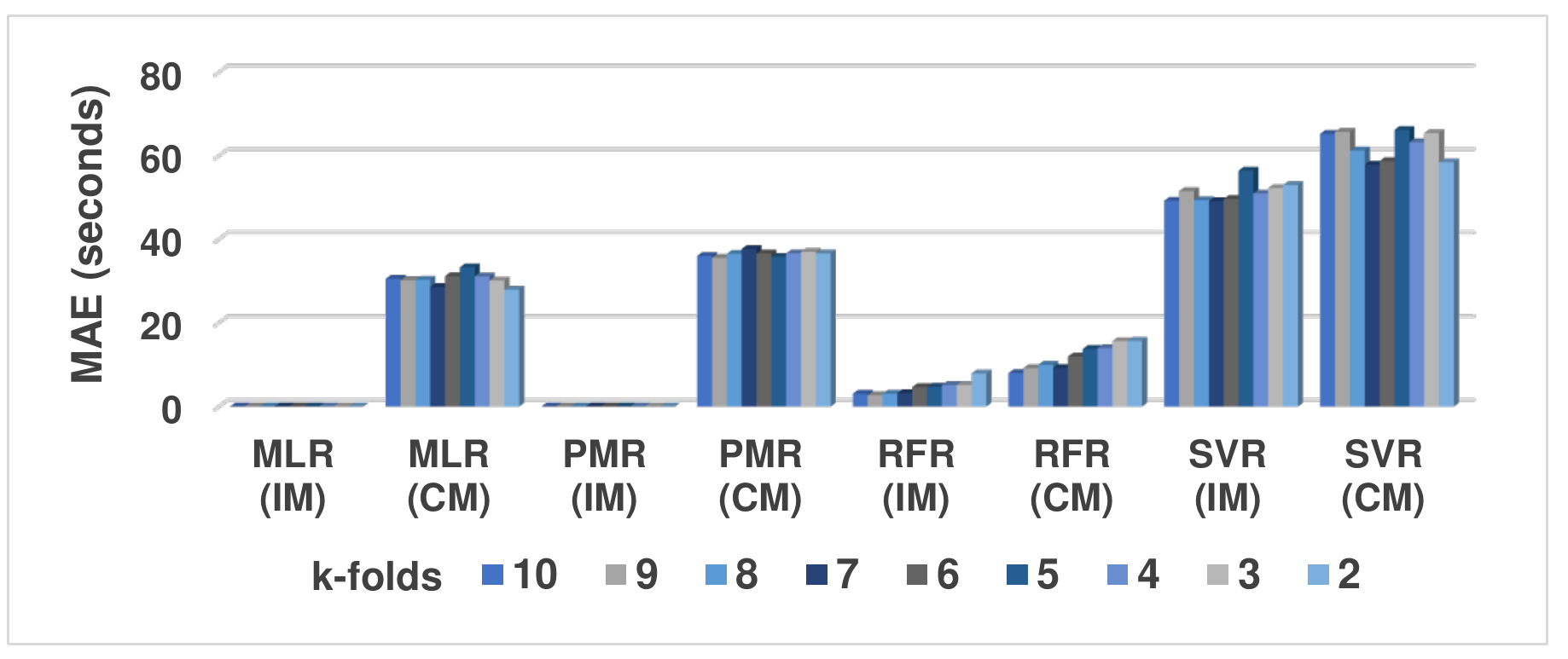}
    \caption{Mean Absolute Error (MAE) of the prediction models for varying sample sizes using k-fold cross validation}
    \label{fig:mae2}
\end{figure}

\textit{Preliminary Results}: 
The goal is to demonstrate the feasibility of estimating the time to offload in the Cloud-to-Fog scenario.
Figure~\ref{fig:accuracy1} shows the accuracy 
for the Collective Model (CM) and Individual Model (IM), using the models discussed in Section~\ref{subsec:models} for varying sizes of training and testing data. PMR and MLR have 100\% accuracy for predicting $t_{offload}$ using IM. 
For the CM, RFR model outperforms other models with an accuracy of 97\%.

Figure~\ref{fig:mae1} shows the Mean Absolute Error (MAE) of the prediction results; a lower value indicates efficient prediction. Predicting $t_{offload}$ from the IM, PMR has the lowest MAE (1.7 seconds), while for the CM, RFR has the lowest MAE (6.76 seconds).

To assess the performance of the prediction models, the models are trained and tested using k-fold cross validation; Figure~\ref{fig:accuracy2} shows the accuracy of the models. The accuracy of the RFR prediction model for the CM, decreases as the number of folds decreases from 96\% to 80\%. While for the Individual Model, MLR and PMR provided the highest accuracy of 100\%.

Figure ~\ref{fig:mae2} shows the MAE of the prediction models for different folds. For the IM, PMR has the lowest MAE (1.06 seconds) and for the CM, RFR has the lowest MAE (3.04 seconds).

The preliminary results indicate that the offloading time can be estimated with reasonable accuracy for container-based service offloading using the Save and Load technique. Using collective models, the random forest regression model yields highest accuracy and for individual models, the multivariate linear and polynomial multivariate regression models yield the highest accuracy.

\section{Related Work}
\label{sec:relatedwork}
Fog offloading is necessary to meet the overall Quality-of-Service requirements of an application. Depending on the offloading scenario the benefits would be for meeting the computational requirements of an application, meeting latency demands, balancing the load, and managing energy consumption~\cite{management2018}. Section~\ref{sec:techniques} highlighted three different offloading scenarios, namely Cloud-to-Fog, Fog-to-Cloud, and Device-to-Fog, which are first considered here and is followed by the approaches used for offloading. 

\textit{Cloud-to-Fog}: Current techniques to partition a monolithic application (if it is not designed as a micro-service based application) are manual~\cite{wang2017enorm}. Multiple approaches have been adopted to make decisions on Fog placement. Examples include heuristics and integer linear programming. A heuristic-based task scheduling algorithm, with the objective to balance between makespan and the monetary cost of Cloud resources has been proposed~\cite{pham2016towards}. A predictive offloading method QCILP (Quadratically Constrained Integer Linear Programming) is developed to minimise the energy costs and meet the latency requirements of services~\cite{chen2018predictive}. 

\textit{Fog-to-Cloud}: Fog resources are anticipated to be hardware constrained and geographically dispersed when compared to a Cloud data center~\cite{management2018}. Therefore, services may not be able to easily scale across the Fog if they have significant compute or storage requirements. Therefore, a Fog service may need to be offloaded back to the Cloud. One strategy adopted to handle the load on the Fog employs an analytical queuing model that is based on the LRU filter~\cite{enguehard2018popularity}.

\textit{Device-to-Fog}: Osmotic computing provides an architecture to deal with Device-to-Fog offloading~\cite{osmoticcomputing-01}. Major concerns that need to be addressed while offloading from user devices or sensors onto the Fog include minimising delays~\cite{yousefpour2018reducing}, the amount of data transferred to the Cloud~\cite{yu2017novel}, and achieving low communication overheads~\cite{hou2019fog}. 

Virtual Machine (VM) and containers are the two main approaches considered for offloading services in the literature.  
VMs when compared to containers have larger overheads and are generally used in resource abundant environments, such as the Cloud. In the context of limited compute and storage resources as seen in the Fog, containers may be more appropriate for offloading services. Nonetheless, VM placement in the Fog by taking user mobility into account using integer linear programming has been proposed~\cite{gonccalves2018proactive}.

The majority of research in Fog computing that focuses on offloading takes containers into account. 
Service hand-off for offloading services between Fog resources is proposed~\cite{ma2017efficient} and live container migration using CRIU (Checkpoint/Restore in Userspace) is considered~\cite{nadgowda2017voyager}.  
To minimise the downtime when a service is offloaded, 
Research that explores the optimisation of the offload process minimises the downtime when a service is offloaded~\cite{machen2016migrating}. 

The above highlighted research presents different techniques that are used to optimise the offloading process or the placement of services in the Fog when a service is offloaded. The research reported in this paper on the other hand aims to characterise the process of offloading, estimate the time taken to offload, and validate models used to estimate the offload time. 

\section{Conclusions}
\label{sec:conclusions}
This paper proposes the design and implementation of methods for estimating the offload time using containers in Fog computing. The `Save and Load' offloading technique and four estimation models, namely Multivariate Linear Regression, Polynomial Multivariate, Random forest and Support Vector Regression are considered. A catalogue of 21 metrics collected at the system and process levels during runtime and offline are used as input to the models. Two estimation methods, namely using a collective model and individual models are proposed. Experimental studies are pursued on two Cloud-Fog platforms and preliminary results indicate that up to 97\% and 100\% accuracy can be obtained using collective models and individual models respectively when estimating the time taken to offload a service from the Cloud to the Fog. 

Future work will explore alternate container-based offloading technique and their effect on the overheads in offloading. 

\begin{acks}
The first author is funded by the Schlumberger Scholarship.
\end{acks}

\bibliographystyle{ACM-Reference-Format}
\bibliography{references.bib}

\end{document}